\documentclass[3p,times,procedia]{elsarticle}
\usepackage{nupha_ecrc}

\volume{00}

\firstpage{1}

\journalname{Nuclear Physics A}

\runauth{}

\jid{nupha}

\jnltitlelogo{Nuclear Physics A}

\usepackage{amssymb}
\usepackage{amsmath}

\usepackage[figuresright]{rotating}

\begin{document}

\begin{frontmatter}



\dochead{XXVIth International Conference on Ultrarelativistic Nucleus-Nucleus Collisions\\ (Quark Matter 2017)}

\title{Universality regained: Kibble-Zurek dynamics, off-equilibrium scaling and the search for the QCD critical point}


\author[BNL]{Swagato Mukherjee}
\author[BNL]{Raju Venugopalan}
\author[MIT]{Yi Yin\footnote{Presenter}} 

\address[BNL]{Department of Physics, Brookhaven National Laboratory, Upton, New York
11973-5000,}
\address[MIT]{Center for Theoretical Physics, Massachusetts Institute of Technology, Cambridge, MA 02139, USA.}

\begin{abstract}
We generalize and apply the key elements of the Kibble-Zurek framework of nonequilibrium phase transitions to study the non-equilibrium critical cumulants near the QCD critical point.
We demonstrate the off-equilibrium critical cumulants are expressible as universal scaling functions.
We discuss how to use off-equilibrium scaling to provide powerful model-independent guidance in searches for the QCD critical point.
\end{abstract}


\end{frontmatter}


\section{Introduction}
\label{intro}

The searches for the conjectured QCD critical point in heavy-ion collision experiments have attracted much theoretical and experimental effort. 
The central question is to identify telltale experimental signals of the presence of the QCD critical point.
In particular, universality arguments can be employed to locate the QCD critical point.

We begin our discussion with the static properties of QCD critical point,
which is argued to lie in the static universality class of the 3d Ising model. 
The equilibrium cumulants of the critical mode satisfy the static scaling relation:
\begin{eqnarray}
\label{eq:kappa-eq}
\kappa^{\rm eq}_{n} \sim \xi_{\rm eq}^{\frac{-\beta+(2-\alpha-\beta)\,(n-1)}{\nu}}\, f^{\rm eq}_{n}(\theta)~\sim \xi_{\rm eq}^{\frac{-1+5\,(n-1)}{2}}\, f^{\rm eq}_{n}(\theta)\, ,
\qquad
n=1,2,3,4\ldots
\end{eqnarray}
where $\xi_{\rm eq}$ is the correlation length, which will grow universally near the critical point and $\theta$ is the scaling variable.
$\alpha, \beta, \nu$ are standard critical exponents. 
Here and hereafter, we will use the approximate value taken from the 3d Ising model: $\alpha\approx 0, \beta\approx 1/3, \nu\approx 2/3$. 
The static scaling relation \eqref{eq:kappa-eq} indicates that non-Gaussian cumulants are more sensitive to the growth of the correlation length (e.g. $\kappa^{\rm eq}_{3}\sim \xi^{4.5}_{\rm eq}, \kappa^{\rm eq}_{4}\sim \xi^{7}_{\rm eq}$ while the Gaussian cumulant $\kappa_{2}\sim \xi^{2}_{\rm eq}$). 
Moreover, for non-Gaussian cumulants, the universal scaling functions $f^{\rm eq}_{n}(\theta)$ can be either positive or negative and depends on $\theta$ only. 
Since critical cumulants contribute to the corresponding cumulants of net baryon number multiplicity, 
the enhanced non-Gaussian fluctuations (of baryon number multiplicities) as well as their change in sign and the associated non-monotonicity as a function of beam energy $\sqrt{s}$ can signal the presence of critical point in the QCD phase diagram. (see Ref.~\cite{Luo:2017faz} for a recent review on experimental measurements.)

However, off-equilibrium effects upset
the naive expectation based on the static properties of the critical system.
The relaxation time of critical modes $\tau_{\rm eff}\sim \xi^{z}$ grows universally with the universal dynamical critical exponent $z\approx 3$ for the QCD critical point~\cite{Son:2004iv}. 
Consequently, critical fluctuations inescapably fall out of equilibrium. 
In Ref.~\cite{Mukherjee:2015swa},  we studied the real time evolution of non-Gaussian cumulants in the QCD critical regime by significantly extending prior work on Gaussian fluctuations~\cite{Berdnikov:1999ph}. 
We found off-equilibrium critical cumulants can differ in both magnitude and sign from equilibrium expectations.
The resulting off-equilibrium evolution of critical fluctuations depend on a number of non-universal inputs such as the details of trajectories in QCD phase diagram and the mapping relation between the QCD phase diagram and the 3d Ising model. 
(See also C.~Herold's, L.~Jiang's, and M.~Nahrgang's talk in this Quark Matter and the plenary talk in the last Quark Matter~\cite{Nahrgang:2016ayr} for related theoretical efforts.). 

Is critical universality completely lost in the complexity of off-equilibrium evolution? 
Are there any universal features of off-equilibrium evolution of critical cumulants?
In Ref.~\cite{Mukherjee:2016kyu}, 
we attempted to answer those questions based on the key ideas of the Kibble-Zurek framework of non-equilibrium phase transitions.
We demonstrated the emergence of off-equilibrium scaling and universality. 
Such off-equilibrium scaling and universality open new possibilities to identify experimental signatures of the critical point. 
The purpose of this proceeding is to illustrate the key idea of Ref.~\cite{Mukherjee:2016kyu} and discuss how to apply this idea to search for QCD critical point. 

\section{Kibble-Zurek dynamics: the basic idea and an illustrative example}
\label{idea}

The basic idea of Kibble-Zurek dynamics was pioneered by Kibble in a cosmological setting and was generalized to describe similar problem in condensed matter system (see Ref.~\cite{Zurek} for a review). 
The Kibble-Zurek dynamics is now considered to be the paradigmatic framework to describe critical behavior out of equilibrium. 
Its application covers an enormous variety of phenomena over a wide range of scales ranging from low-temperature physics to astrophysics . 
We will to use this idea to explore the dynamics of QCD matter near the critical point. 

\begin{figure*}[!hbt]
\begin{center} \vspace{-0.1in}
\includegraphics[width=0.32\textwidth]{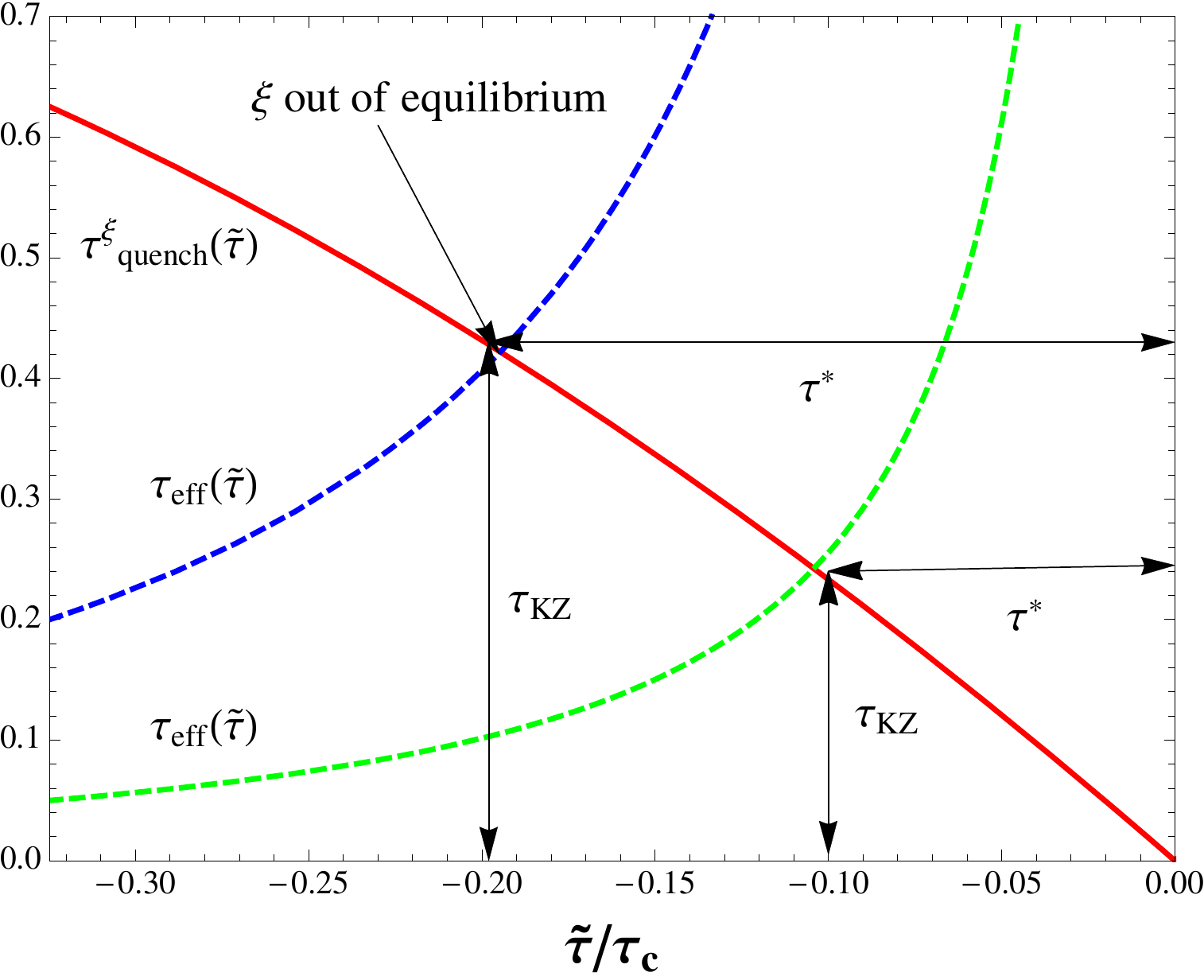} \hspace{0.022in}
\includegraphics[width=0.32\textwidth]{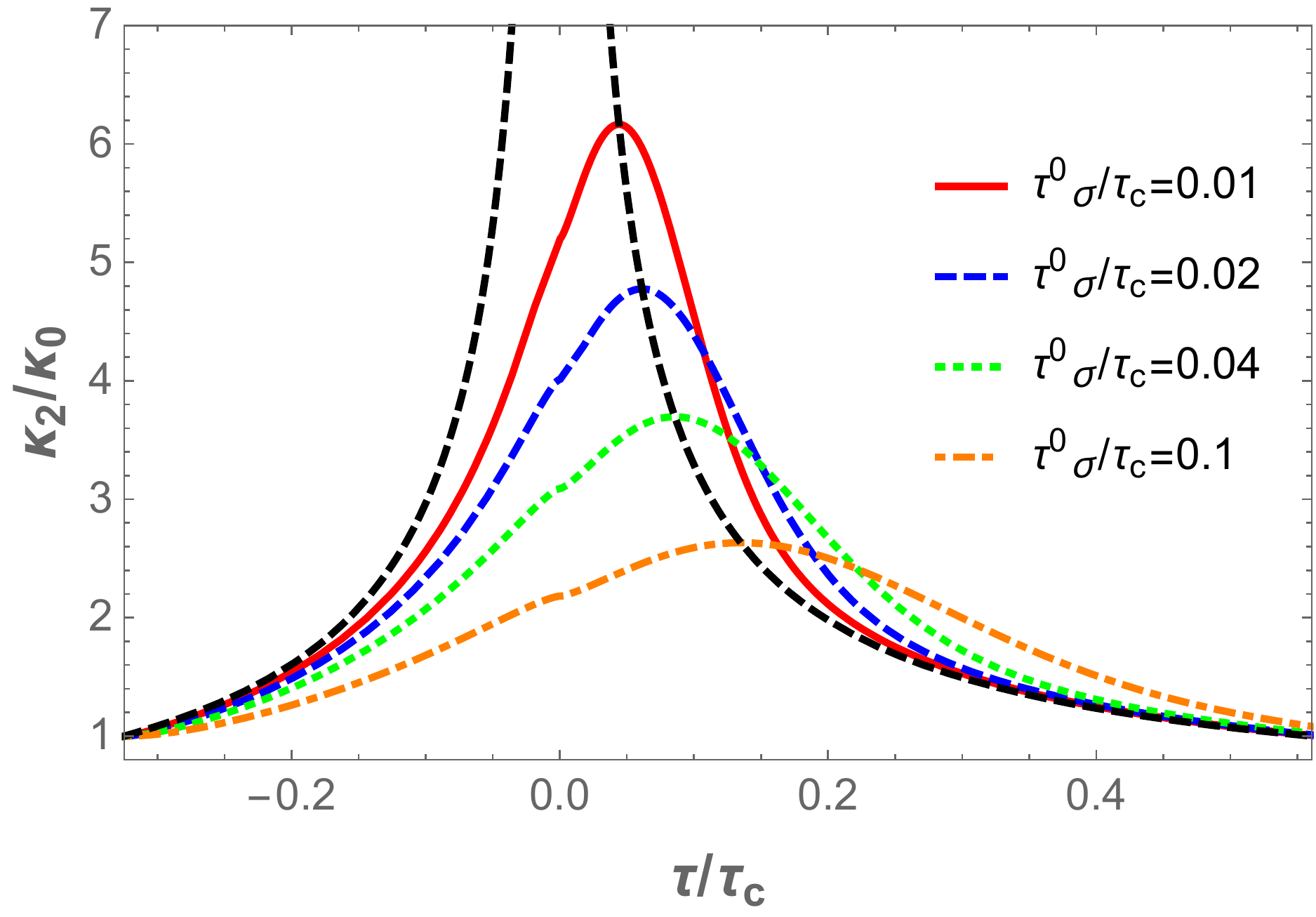} \hspace{0.022in}
\includegraphics[width=0.32\textwidth]{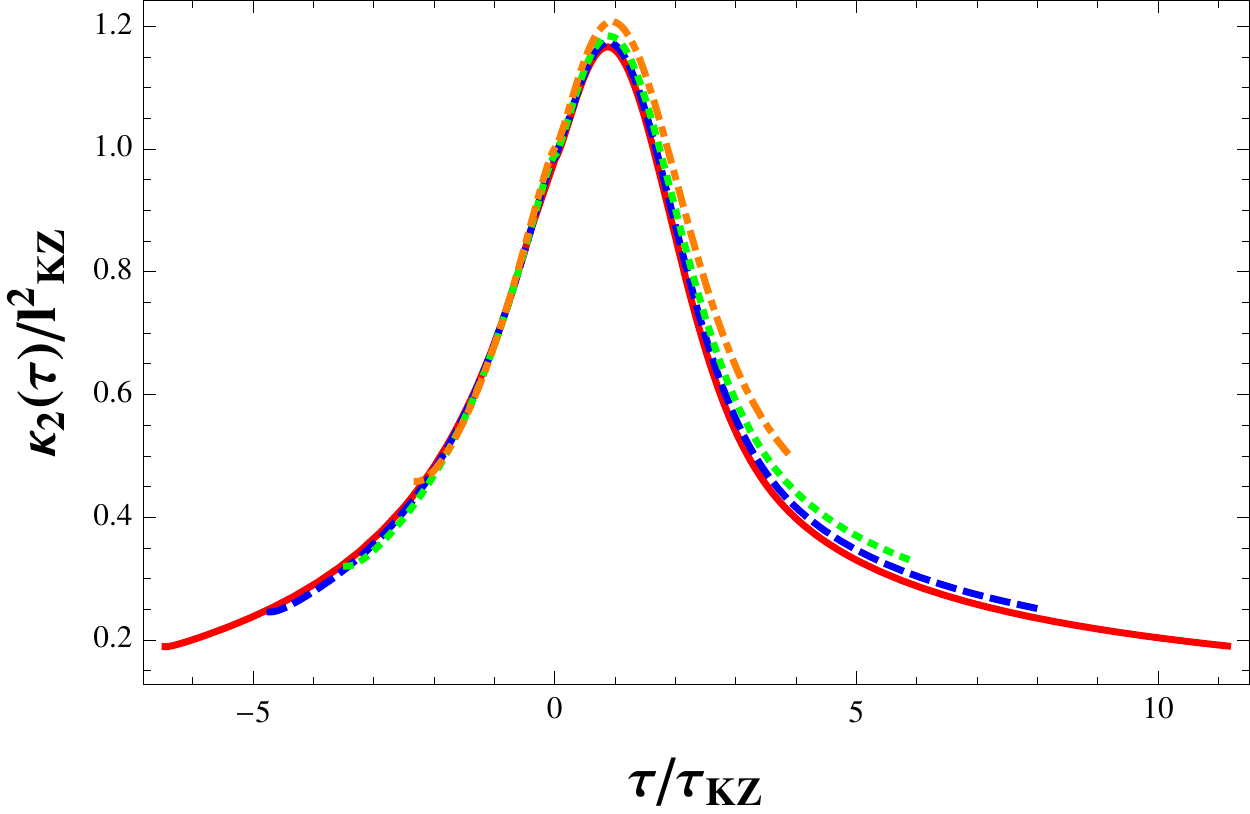}
\vspace{-0.15in}
\caption{(color online) 
(left):
the representative temporal evolution of the relaxation time $\tau_{\rm eff}$ (blue and green dashed curve) and quench time $\tau_{\rm quench}$ (red solid curve) near the critical point.  
$\tau$ is the proper time parametrizing the trajectory and is chosen to be $0$ when system is at the critical point. (middle): The off-equilibrium evolution of Gaussian cumulants $\kappa_{2}$ determined by solving rate equation proposed in Ref.~\cite{Berdnikov:1999ph}. 
Results with different non-universal inputs are shown in red, dashed blue, dotted green and dot-dashed orange curves respectively.
The dashed black curve plots the equilibrium value. 
(right): the rescaled  Gaussian cumulants $\kappa_{2}/l^{2}_{\rm KZ}$ vs  the rescaled time $\tau/\tau_{\rm KZ}$. }  
\label{fig:KZ-idea}
\end{center}
\vspace{-0.2in}
\end{figure*} 

%
%
%

The dramatic change in the behavior of the quench time relative to the relaxation time is at the heart of the KZ dynamics.
As the system approaches the critical point, the relaxation time grows due to the critical slowing down (c.f. dashed curves in ~Fig.~\ref{fig:KZ-idea} (left)). 
In contrast, 
the quench time $\tau_{\rm quench}$, defined as the relative change time of correlation length, becomes shorter and shorter due to the rapid growth of the correlation length (c.f.~Fig.~\ref{fig:KZ-idea} (left)) . 
Therefore at some proper time, say $\tau^{*}$, $\tau_{\rm eff}$ is equal to $\tau_{\rm quench}$. 
After $\tau^{*}$,
the system expands too quickly (i.e. $\tau_{\rm quench} > \tau_{\rm eff}$) for the evolution of correlation length to follow the growth of equilibrium correlation length. 
It is then natural to define an emergent length scale, the Kibble-Zurek length $l_{\rm KZ}$, 
which is the value of the equilibrium correlation length at $\tau^{*}$. 
In another word,
$l_{KZ}$ is the maximum correlation length the system could develop when passing through the critical point. 
Likewise, 
one could introduce an emergent time scale, referred as Kibble-Zurek time, $\tau_{\rm KZ}=\tau_{\rm eff}(\tau^{*})$, 
which is the relaxation time at $\tau^{*}$. 
If the critical fluctuations are in thermal equilibrium, their magnitudes are completely determined by the equilibrium correlation length. 
For off-equilibrium evolution, we expect that magnitude of off-equilibrium fluctuations is controlled by $l_{\rm KZ}$ and their temporal evolutions are characterized by $\tau_{\rm KZ}$.  

The emergence of such a characteristic length scale $l_{\rm KZ}$ and the time scale $\tau_{\rm KZ}$ indicates that cumulants will scale as a function of them. 
As an illustrative example, 
we revisited the study of Ref.~\cite{Berdnikov:1999ph}. 
Fig.~\ref{fig:KZ-idea} (middle) plots the evolution of the off-equilibrium Gaussian cumulant $\kappa_{2}$ by solving the rate equation proposed in Ref.~\cite{Berdnikov:1999ph} from different choices of non-universal inputs. 
Those off-equilibrium cumulants are different from the equilibrium one and look different at first glance. 
What happens if we rescale those Gaussian cumulants by $l^{2}_{\rm KZ}$ and present their temporal evolution as a function of the rescaled time $\tau/\tau_{\rm KZ}$?
As shown in Fig.~\ref{fig:KZ-idea} (right),
the rescaled evolutions now look nearly identical, which confirms the expectation from the KZ scaling.
\section{Off-equilibrium scaling of critical cumulants}
\label{result}

In Ref.~\cite{Mukherjee:2016kyu},
we proposed the following scaling hypothesis for the off-equilibrium evolution of critical cumulants:
\begin{eqnarray}
\label{eq:scaling}
\kappa_{n}\left(\tau;\Gamma\right) \sim l^{\frac{-1+5\,(n-1)}{2}}_{\rm KZ}\, \bar{f}_{n}\left(\,\frac{\tau}{\tau_{\rm KZ}}, \theta_{\rm KZ}\,\right)\, , 
\qquad
n=1,2, 3, 4\ldots ,
\end{eqnarray}
which is motivated in part by the equilibrium scaling relation \eqref{eq:kappa-eq} and in part by the key ideas of Kibble-Zurek dynamics. 
The scaling hypothesis \eqref{eq:scaling} extends the scaling hypothesis for Gaussian cumulants~(see for example Ref.~\cite{Gubser}) to non-Gaussian cumulants, which have thus far received little attention in the literature. 
We further introduced the Kibble-Zurek scaling variable $\theta_{\rm KZ}$. 
In analogue to $l_{\rm KZ}$ and $\tau_{\rm KZ}$, 
$\theta_{\rm KZ}$ is defined as the value of the scaling variable $\theta$ when the quench time is equal to the relaxation time and can be considered as the memory of the sign of non-Gaussian cumulants of the system. 
The off-equilibrium scaling functions $\bar{f}_{n}$ in \eqref{eq:scaling} is universal and does not explicitly depend on non-universal inputs which are collectively denoted by $\Gamma$. 
The dependence of those off-equilibrium cumulants $\kappa_{n}(\tau;\Gamma)$ on non-universal inputs $\Gamma$ are absorbed in $\tau_{\rm KZ}(\Gamma), l_{\rm KZ}(\Gamma)$ and $\theta_{\rm KZ}(\Gamma)$.

\begin{figure*}[!hbt]
\begin{center} \vspace{-0.1in}
\includegraphics[width=0.32\textwidth]{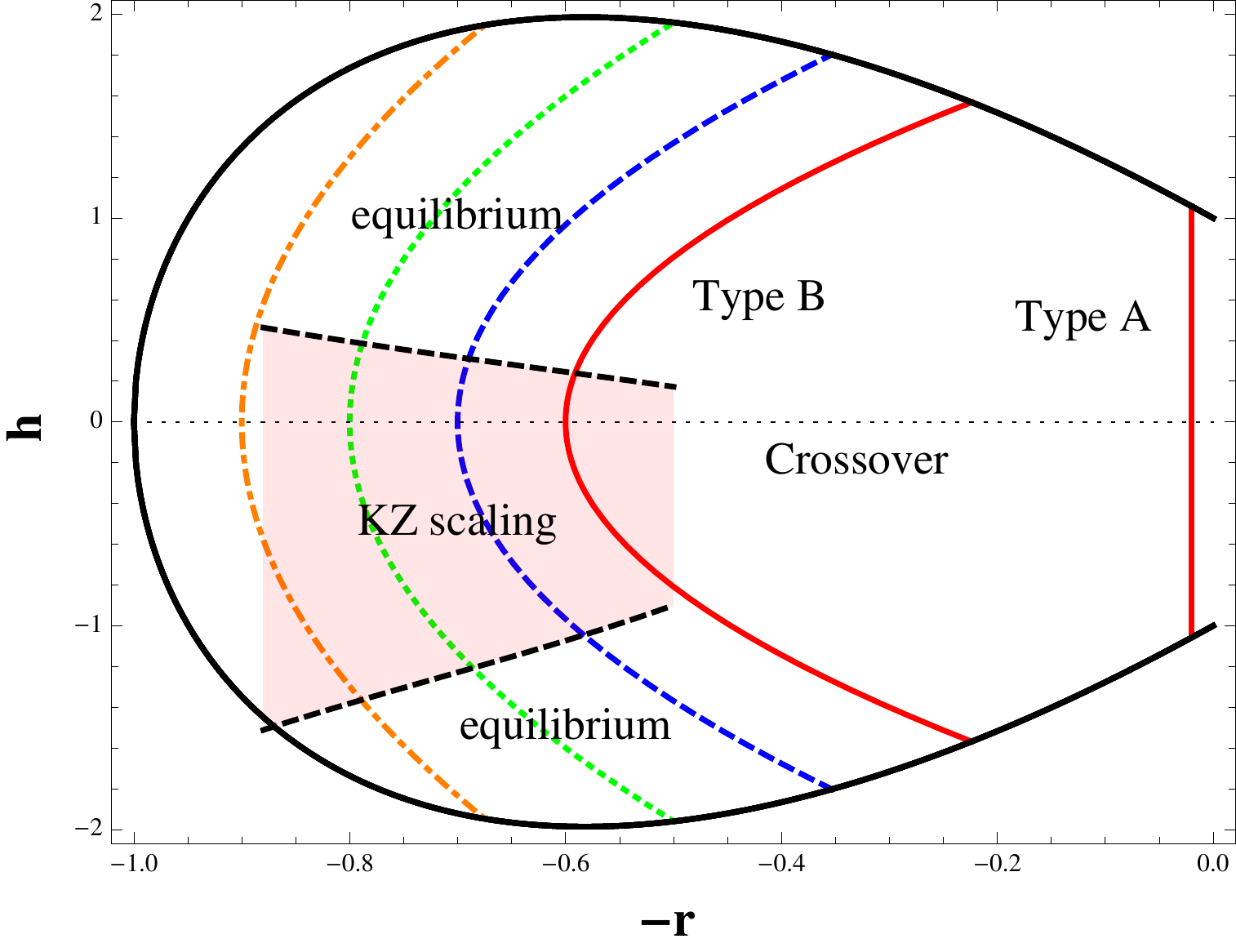} \hspace{0.025in}
\includegraphics[width=0.32\textwidth]{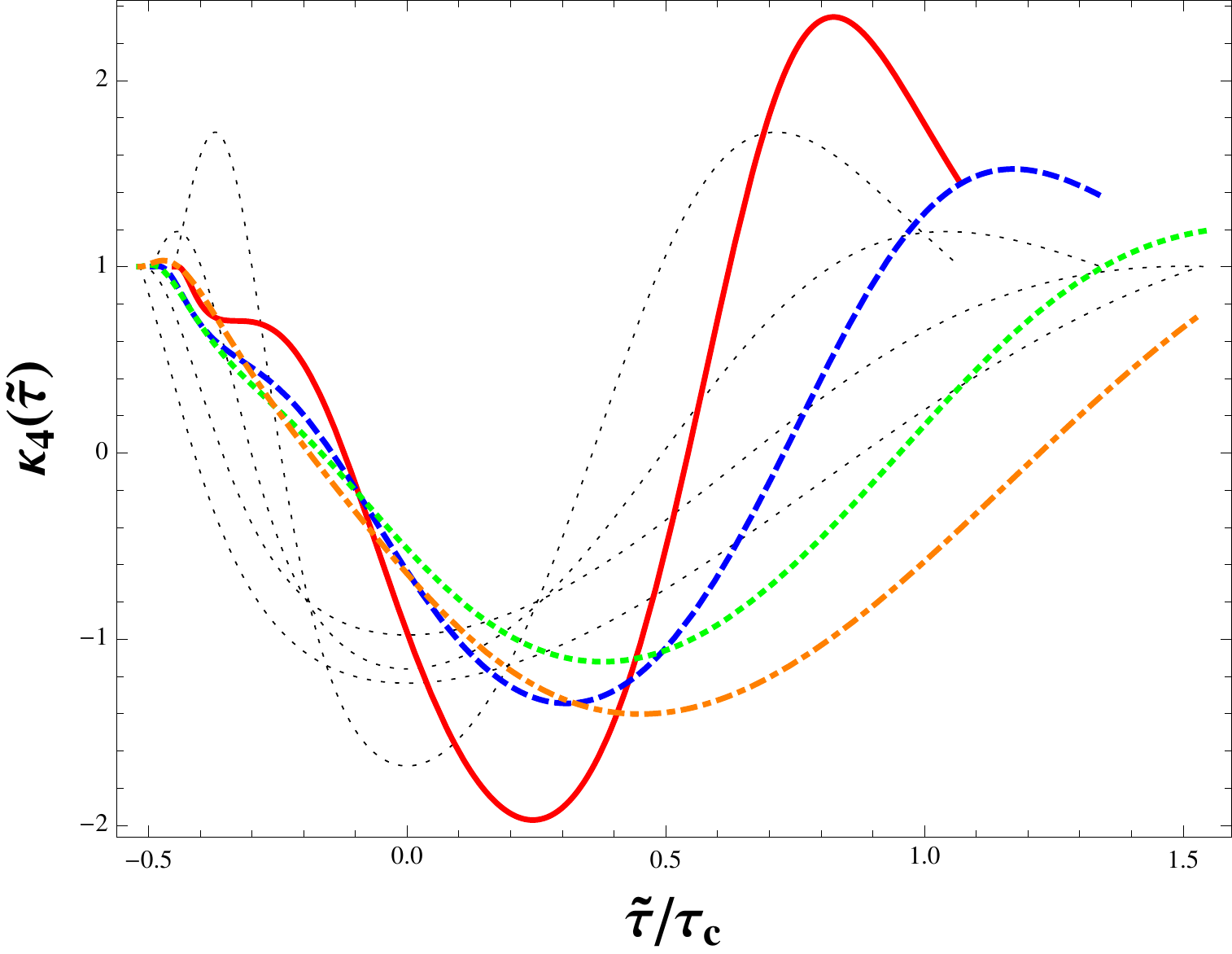} \hspace{0.025in}
\includegraphics[width=0.32\textwidth]{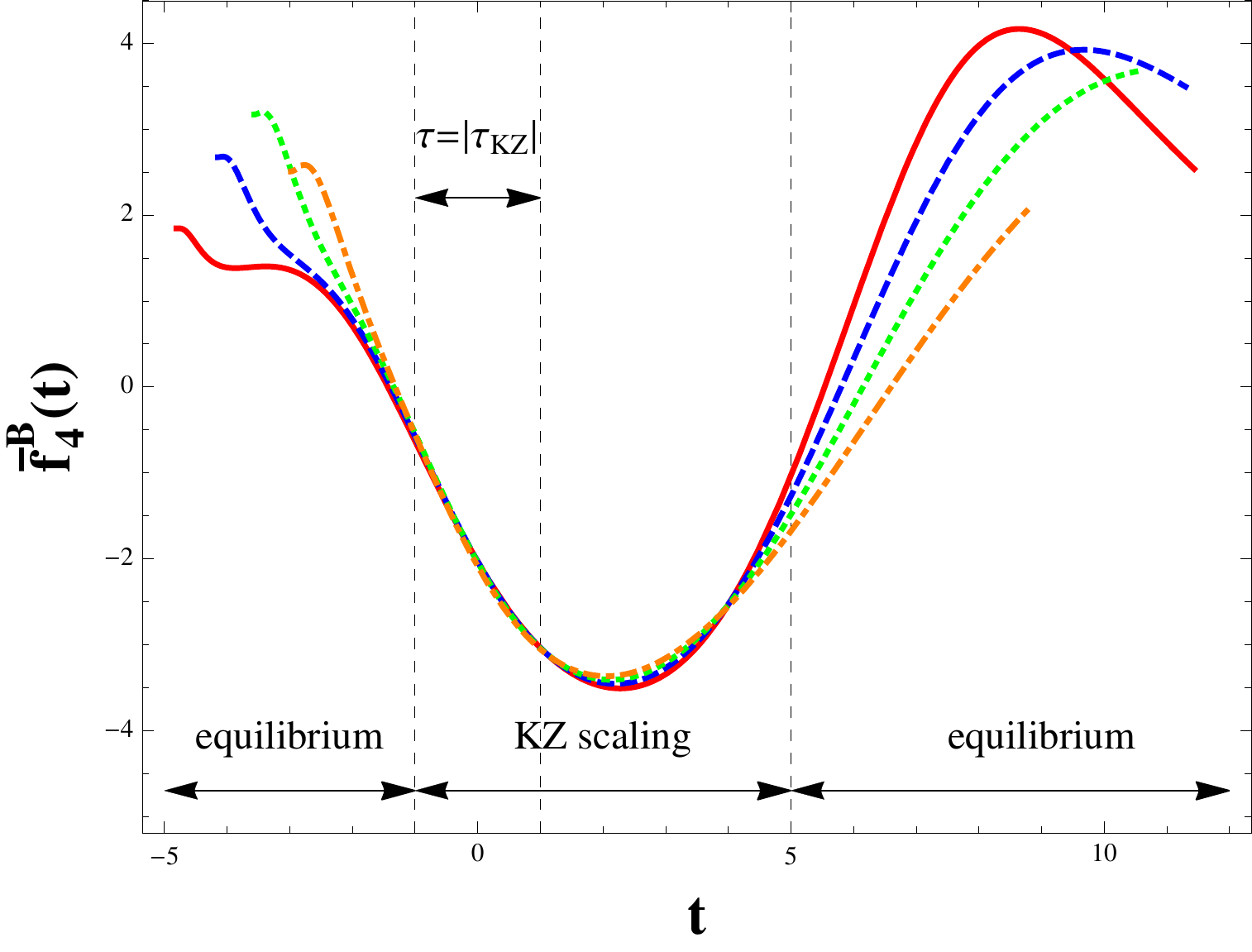}
\vspace{-0.15in}
\caption{(color online) 
(Left):
Sketch of trajectories on the crossover side of the critical point. 
The solid curve delineates the boundary of the critical regime in the $r-h$ plane where $r$ is the reduced Ising temperature and $h$ is rescaled Ising magnetic field. 
A possible KZ scaling regime is illustrated in the shaded area.
 (Middle):
 Nonequilibrium evolution of the kurtosis $\kappa_{4}$ (normalized
by its initial equilibrium value) for representative trajectories. 
 The corresponding equilibrium values are plotted in dotted curves. 
 (Right): 
 The rescaled kurtosis vs the rescaled time $t=\tau/\tau_{\rm KZ}$. 
 We choose $\tau$ in such a way that it becomes $0$ when the system reaches the crossover curve. 
} 
 \label{fig:scaling}
\end{center}
\vspace{-0.2in}
\end{figure*} 

Since $\bar{f}_{n}(\tau/\tau_{\rm KZ}, \theta_{\rm KZ})$ in \eqref{eq:scaling} is a universal scaling function, 
the temporal evolution of rescaled cumulants would become similar if those evolutions share the same $\theta_{\rm KZ}$. 
In Fig.~\ref{fig:scaling} (middle), 
we present the evolution of the fourth cumulants, kurtosis $\kappa_{4}$ from four different trajectories (c.f.~ Fig.~\ref{fig:scaling} (left)).
Those evolutions are determined from different non-universal inputs. 
We however combine those non-universal inputs in such a way that the resulting $\theta_{\rm KZ}$ are the same. 
The rescaled kurtosis vs $\tau/\tau_{\rm KZ}$ is plotted in Fig.~\ref{fig:scaling} (right).
Those rescaled kurtosis look different away from the crossover curve since they are close to their corresponding equilibrium which are different for different trajectories. 
However, 
in the vicinity of the crossover curve, 
those rescaled kurtosis again merge into a single curve, confirming the validity of our hypothesis.  
For further numerically test of the scaling hypothesis as well as analytical insights into scaling form, 
see Ref.~\cite{Mukherjee:2016kyu}.

\section{The off-equilibrium scaling as a signature of the QCD critical point}
\label{summary}

We now consider the implications of our findings for the beam energy scan (BES) search for the QCD critical point.
The critical contribution to the cumulants of baryon number multiplicity $\kappa^{\rm data}_{n}$ measured in experiment is proportional to the critical cumulants on the freeze-out curve. 
The scaling hypothesis \eqref{eq:scaling} indicates:
\begin{eqnarray}
\label{eq:scalingf}
\tilde{\kappa}^{\rm data}_{n}\equiv\frac{\kappa^{\rm data}_{n}\left(\tau_{f};\Gamma\right)}{l^{-\frac{1}{2}+\frac{5}{2}(n-1)}_{\rm KZ}}\propto\frac{\kappa_{n}\left(\tau_{f};\Gamma\right)}{l^{-\frac{1}{2}+\frac{5}{2}(n-1)}_{\rm KZ}} \sim \, \bar{f}_{n}\left(\,\frac{\tau_{f}}{\tau_{\rm KZ}}, \theta_{\rm KZ}\,\right)\, . 
\end{eqnarray}
Note $\tau_{f}$ is the proper time of the system on the freeze-out curve and the universal scaling function $\bar{f}_{n}$ does not depend on the non-universal inputs collectively denoted by $\Gamma$.

The scaling relation \eqref{eq:scalingf} means that the rescaled data $\tilde{\kappa}^{\rm data}_{n}$ will only be sensitive to $\left(\tau_{f}/\tau_{\rm KZ}, \theta_{\rm KZ}\right)$. 
In another word, 
let us rescale an ensemble of experimental data in heavy-ion collisions with different beam energy $\sqrt{s}$, impact parameter $b$ (or centrality) and rapidity $\eta$ using the appropriate power of $l_{KZ}$ according to \eqref{eq:scalingf}, 
those collisions with the similar $\left(\tau_{f}/\tau_{\rm KZ}, \theta_{\rm KZ}\right)$ will have approximately the similar values for rescaled data if the QCD critical regime is probed by those collisions.  

A crucial step in testing scaling as outlined above is to determine $l_{KZ}, \tau_{KZ},\theta_{KZ}$ and the freeze-out time $\tau_{f}$ as a function of $\left(\sqrt{s}, b, \eta\, \right)$. 
We need to use hydrodynamical simulations to determine trajectories in the phase diagram as well as the expansion rate of the system for a given collision of $\left(\sqrt{s}, b, \eta\,\right)$. 
We then compare the relaxation time and quench time to determine the corresponding  $l_{KZ}\left(\sqrt{s}, b, \eta\, \right)$, $\tau_{KZ}\left(\sqrt{s}, b, \eta\,\right)$, $\theta_{KZ}\left(\sqrt{s}, b, \eta\,\right)$. 
Since those scaling variables are non-trivial functions of $\left(\sqrt{s}, b, \eta\, \right)$, 
the success of such theory-data comparisons would provide strong evidence for the existence of the QCD critical point. 
The study of procedures outlined above, with examples including mock BES data is in progress~\cite{search}. 
\\

 {\bf Acknowledgments.} 
This material is based upon work supported by the U.S. Department of Energy, Office of Science, Office of Nuclear Physics, under Contract Number DE-SC0012704 (SM, RV), DE-SC0011090 (YY) and within the framework of the Beam Energy Scan Theory (BEST) Topical Collaboration.





\bibliographystyle{elsarticle-num}



\end{document}